\documentclass[twocolumn,amsmath,amssymb,aps,prl,showpacs,tightenlines]{revtex4}
\usepackage{amsmath}
\usepackage{amssymb}
\usepackage{txfonts}
\usepackage{graphicx}
\usepackage{float}
\usepackage{epstopdf}
\usepackage{dcolumn}
\usepackage{bm}
\usepackage[colorlinks=true,citecolor=blue,linkcolor=magenta,urlcolor=blue]{hyperref}
\usepackage{indentfirst}
\begin{document}

\title{Ultra-low threshold chaos in cavity magnomechanics}

\author{Jiao Peng$^{1,2}$}
\author{Zeng-Xing Liu$^{1}$}\email{zengxingliu@hust.edu.cn}
\author{Ya-Fei Yu$^{2}$}\email{20031115@m.scnu.edu.cn}
\author{Hao Xiong$^{3}$}

\affiliation{$^{1}$School of Electronic Engineering $\&$ Intelligentization, Dongguan University of Technology, Dongguan, Guangdong 523808, China}
\affiliation{$^{2}$School of Information and Optoelectronic Science and Engineering, South China Normal University, Guangzhou, Guangdong 510006, China}
\affiliation{$^{3}$School of physics, Huazhong University of Science and Technology, Wuhan 430074, China}

\date{\today}

\begin{abstract}
Cavity magnomechanics using mechanical degrees of freedom in ferromagnetic crystals provides a powerful platform for observing many interesting classical and quantum nonlinear phenomena in the emerging field of magnon spintronics.
However, to date, the generation and control of chaotic motion in a  cavity magnomechanical system remain an outstanding challenge due to the inherently weak nonlinear interaction of magnons.
Here, we present an efficient mechanism for achieving magnomechanical chaos, in which the magnomechanical system is coherently driven by a two-tone microwave field consisting of a pump field and a probe field.
Numerical simulations show that the relative phase of the two input fields plays an important role in controlling the appearance of chaotic motion and, more importantly, the threshold power of chaos is reduced by 6 orders of magnitude from watts (W) to microwatts ($\mu$W).
In addition to providing insight into magnonics nonlinearity, cavity magnomechanical chaos will always be of interest because of its significance both in fundamental physics and potential applications ranging from ultra-low threshold chaotic motion to chaos-based secret information processing.
\end{abstract}

\maketitle

\section{INTRODUCTION}

Cavity magnomechanics is a rapidly developing research field that provides a special platform for observing many interesting classical and quantum phenomena \cite{A. V. Chumak2015,Hybrid,H.Y. Yuan2022,S. Zheng2023}. In a magnomechanical system, the ferromagnetic Kittel mode (a uniform mode of spin waves) of the Yttrium Iron Garnet (YIG) sphere \cite{YIG} can couple to the mechanical degrees of freedom via radiation pressure-like magnetostrictive interaction \cite{J. Holanda2018,magnomechanics4,M. Yu2020,Magnon1,Magnon2,Y. Xu2021,C. S. Zhao2022,J. Li2018,Y. T. Chen2021,x.-L. Hei2023,W. Qiu2022,B. Hussain2022,J. Li2019,Zhang W2021,Squeezing,T. X. Lu2023,C. Kong2019,G.-T. Xu2023} (also known as magnetostrictive effect \cite{E.G. Spencer1958,E.G. Spencer1970}).
Experimental manipulation of the Kittel mode and vibrational mode via magnetostrictive effects has been demonstrated experimentally \cite{J. Holanda2018,magnomechanics4,M. Yu2020,Magnon1,Magnon2,x.-L. Hei2023}, and many intriguing phenomena have been reported in cavity magnomechanics, ranging from magnomechanically induced transparency \cite{magnomechanics4,C. S. Zhao2022} and magnetostrictive-induced slow-light effect \cite{T. X. Lu2023,C. Kong2019} to entanglement and squeezing states of
magnons \cite{J. Li2018,Y. T. Chen2021,W. Qiu2022,B. Hussain2022,J. Li2019,Zhang W2021,Squeezing}
and the ground-state cooling of mechanical vibration
mode \cite{A. Kani2022,Z.-X. Yang2020,Z. Yang2023,M. Asjad2023}. These effects are similar to those obtained via the mechanical effects of light in cavity optomechanics \cite{optomechanics,optomechanics1,optomechanics2}, opening a new way for providing a new type of matter-matter interaction based on the mechanical effects of magnons.

In the past few years, a large number of studies have shown that cavity magnomechanical systems exhibit rich but extraordinary nonlinear effects \cite{kerr11,kerr2,kerr3,comb1,comb2,comb3,comb33,comb4,kerr,Magnomechanics1,Magnomechanics2,Magnomechanics3}.
Recently, an experiment has showed that three different kinds of nonlinearities can be simultaneously activated under a strong microwave drive field, namely, magnetostriction, magnon self-Kerr, and magnon-phonon cross-Kerr nonlinearities, and the Kerr-modified mechanical bistability has been observed \cite{kerr3}.
Furthermore, the generation of magnonic frequency combs based on the resonantly enhanced magnetostrictive effect is predicted theoretically \cite{comb1,comb2,comb3} and quickly verified experimentally \cite{comb4}.
However, as a kind of nonlinear motion prevalent in nature, the generation and manipulation of chaos \cite{chaos10} based on the mechanical effect of magnon is still a prominent challenge due to the weak nonlinear interaction of magnons \cite{comb2,comb3,comb33}.
The studying of cavity magnomechanical chaos, undeniably, is one of the most important aspects of exploring nonlinear properties in cavity magnomechanics \cite{Magnomechanics1,Magnomechanics2,Magnomechanics3}.
In additional, the investigation of magnomechanical chaos may provide theoretical support for the realization of chaos-based secret information processes and quantum communication in the field of magnonics \cite{chaos9,A. Argyris2005,A. B. Ustinov2021,G. D. Vanwiggeren1998}.

In the present work, we propose an effective mechanism for realizing magnomechanical chaos by introducing phase modulation.
The system is coherently driven by a two-tone microwave field consisting of a pump field and a probe field, where the relative phase of the two input fields plays an important role in controlling
the appearance of chaotic motion and the corresponding chaotic dynamics.
With state-of-the-art experimental parameters \cite{magnomechanics4,kerr3}, we show that the threshold power of chaotic motion is significantly reduced by six orders of magnitude, which effectively solves the bottleneck that the weak magnetostrictive interaction cannot trigger chaotic motion in the cavity magnomechanical system.
\begin{figure}[htbp]
\centering
\includegraphics [width=1.0\linewidth] {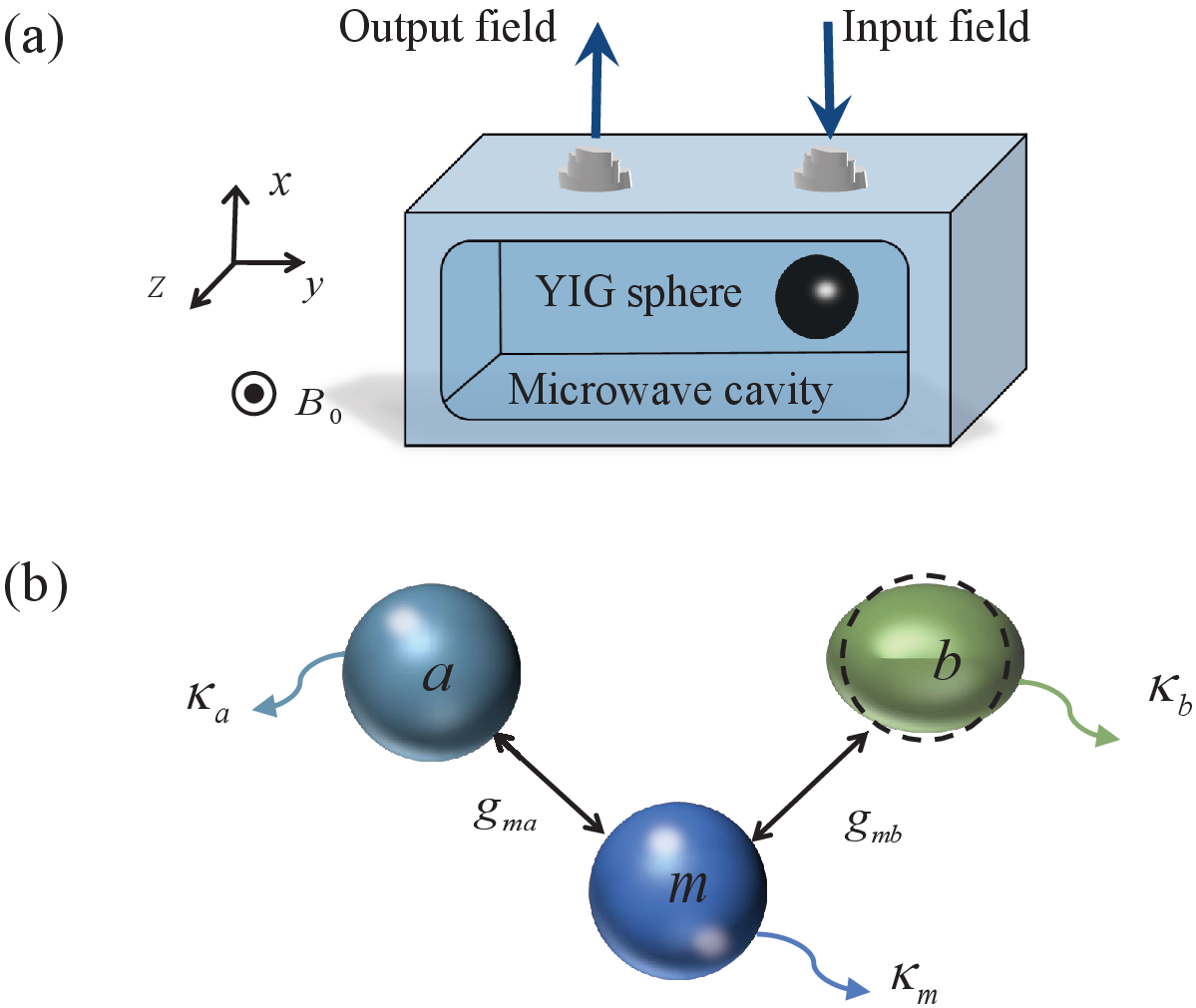}
\caption{(a) Schematic illustration of the cavity magnomechanical system, in which a highly polished YIG sphere is mounted in a three-dimensional microwave cavity.
A two-tone microwave drive field is used to excite the ferromagnetic Kittel mode (magnon mode) of the YIG sphere.
A uniform magnetic field (with the strength $B_0$) along the z direction is applied to saturate the magnetization, and the YIG sphere will be deformed in response to the external magnetic field.
(b) Schematic diagram of the coupling (with the coupling strength $g_{ma}$ and $g_{mb}$) between the microwave cavity mode, the magnon mode, and the deformation mode (vibrational mode) with decay rates $\kappa_{a}$, $\kappa_{m}$, and $\kappa_{b}$, respectively.}
\label{fig:1}
\end{figure}
Furthermore, the influence of the inherent magnon Kerr nonlinearity \cite{kerr11,kerr3,kerr} on chaotic dynamics is also discussed in detail, and the results suggest that the Kerr coefficient plays an important role in the chaotic degree of the system.
Our scheme provides a new perspective for the study of chaotic behavior of magnons and suggests that cavity magnomechanics with inherent nonlinearity is a good platform to explore chaotic phenomena by introducing phase modulation.

\section{Physical model and methods}

The physical model we consider is a cavity magnomechanical system, as schematically shown in Fig. \ref{fig:1}(a), in which a highly polished YIG sphere is placed in a three-dimensional microwave cavity \cite{magnomechanics4}.
The microwave drive field is introduced into the microwave cavity through the input port, and the ferromagnetic Kittel mode (magnon mode) of the YIG sphere will be excited \cite{ferromagnetic}.
Furthermore, a uniform static bias magnetic field (with the strength $B_0$) is applied to the YIG sphere to saturate the magnetization and establish the coupling between the magnon mode and the microwave mode \cite{Strong}.
As shown in Fig. \ref{fig:1}(b), the magnon mode coupled to the microwave cavity mode through the magnetic dipole interaction with the coupling strength $g_{ma}$.
The frequency of the magnon mode ${\omega}_m$ is directly proportional to the strength of the bias magnetic field, i.e., $\omega _m  = \gamma B_0$ with the gyromagnetic ratio $ \gamma /2\pi  = 28\ \rm GHz/\rm T$ \cite{kerr}.
According to the magnetostriction effect \cite{E.G. Spencer1958,E.G. Spencer1970,ferromagnetic2}, the different magnetization induced by the magnon excitation will cause the deformation of the YIG sphere, and at the same time, the deformation of the YIG sphere in response to the external magnetic field can also impact on the magnetization, which gives rise to the coupling between the magnon mode and the vibrational mode \cite{magnomechanics4,Magnon1,Magnon2}.
As shown in Fig. \ref{fig:1}(b), the magnetostrictive force leads to the coupling between deformation and magnetostatic modes with the coupling strength $g_{mb}$.
The magnetostrictive interaction can be described by a radiation pressure-like Hamiltonian, i.e., $\hat H_{int} = \hbar g_{mb} \hat m^\dag  \hat m(\hat b + \hat b^\dag )$, where $\hbar$ is the reduced Planck's constant and $\hat b(\hat b^\dag)$ is the boson annihilation (creation) operator of the deformation mode \cite{magnomechanics4}. $\hat{m}=\sqrt{\frac{V_{m}}{2\hbar\varrho {\rm{M}}}}(M_{x}-iM_{y})$ is the annihilation operator of the magnon mode, with $V_{m}$ the YIG sphere volume, ${\rm{M}}$ the saturation magnetization, and $M_{x,y,z}$ the magnetization components \cite{kerr2}.
Furthermore, we assume that the system is driven by a two-tone microwave driving field consisting of a pump field with the central frequency $\omega _d$, the pump power ${\rm{P}} _d$, the driving amplitude $\varepsilon _d  = \sqrt {{\rm{P}}_d /(\hbar \omega _d )}$, the initial phase $\varphi_{d}$, and a probe field with the central frequency $\omega _p$, the pump power ${\rm{P}} _p$, the driving amplitude $\varepsilon _p  = \sqrt {{\rm{P}}_p /(\hbar \omega _p)}$, the initial phase $\varphi_{p}$, respectively.
Therefore, the Hamiltonian of such cavity magnomechanical system can be written as
\begin{eqnarray}\label{eqn:1}
  \hat H &=& \hbar \omega _a \hat a^\dag  \hat a + \hbar \omega _m \hat m^\dag  \hat m + \hbar \omega _b \hat b^\dag  \hat b + \hbar g_{ma} (\hat a^\dag  \hat m + \hat a \hat m^\dag  )\nonumber \\
  &+&\hbar g_{mb} \hat m^\dag  \hat m(\hat b + \hat b^\dag  )+ {\rm{K}}_m \hat m^\dag  \hat m \hat m^\dag  \hat m \nonumber \\
  &+&\hbar \sqrt {{\kappa _{\rm{1}}}} \big\{{\varepsilon _d}\big[a{e^{i({\omega_{d}t+\varphi _d})}} + {a^\dag }{e^{-i({\omega_{d}t+\varphi _d})}}\big] \nonumber \\
  &+&{\varepsilon_p}\big[a{e^{i({\omega_{p}t+\varphi _p})}} + {a^\dag }{e^{-i({\omega_{p}t+\varphi _p})}}\big]\big\},
\end{eqnarray}
where $\hat a$ and $\hat a^\dag$ are the annihilation and creation operators of the microwave cavity mode with the intrinsic frequency $\omega _a$.
$\omega _m$ and $\omega _b$ are the intrinsic frequencies of the ferromagnetic Kittel mode and the vibrational mode respectively.
$\kappa_1$ refers to the loss rate of the microwave cavity mode associated with the input coupling.
It is worth noting that the YIG sphere also possesses an intrinsic magnon Kerr nonlinearity due to the magnetocrystalline anisotropy \cite{kerr11,kerr2,kerr3,kerr}. Taking the intrinsic magnon Kerr nonlinearity into account, i.e., ${\rm{K}}_m \hat m^\dag  \hat m \hat m^\dag  \hat m$, where ${\rm{K}}_{\rm{m}} {\rm{  = }}\mu _{\rm{0}} {\rm{K}}_{an} \gamma ^{\rm{2}} {\rm{/(M}}^{\rm{2}} {\rm{V}}_{\rm{m}} {\rm{)}}$ is the Kerr nonlinear coefficient, with the vacuum permeability ${\mu}_0$, the first-order magnetocrystalline anisotropy constant ${\rm{K}}_{an}$, and the gyromagnetic ratio ${\gamma}$ \cite{kerr2}.
Note that the Kerr coefficient can be positive or negative depending on which crystallographic axis [100] or [110] of the YIG sphere is aligned in the direction of the static magnetic field $B_{o}$ \cite{kerr}.
It should be pointed out that under a strong microwave drive field, three kinds of nonlinearity, i.e., magnetostriction, magnon self-Kerr, and magnon-phonon cross-Kerr nonlinearities can be simultaneously activated in the cavity magnomechanical system \cite{kerr3}.
However, the cross Kerr coefficient is three orders of magnitude smaller than the self-Kerr coefficient \cite{kerr,kerr3}, so the effect of cross Kerr nonlinearity on chaotic motion is not included in our model.

The dynamics of the magnomechanical system can be described by the Heisenberg-Langevin equations, and thus, in a frame rotating with the microwave drive frequency ${\omega}_d$, we can obtain that
\begin{eqnarray}\label{eqn:2}
\begin{aligned}
 \dot a = & (- i\Delta_a-\frac{{\kappa _a }}{2})a-ig_{ma} m-i\sqrt{\kappa_1}\big[\varepsilon_d e^{-i\varphi_d}+\varepsilon_p e^{-i(\Delta_{p}t+\varphi_p)}\big], \\
 \dot b = & (- i\omega_b-\frac{{\kappa _b }}{2})b-ig_{mb} m^\dag  m , \\
 \dot m = & (- i\Delta_m-\frac{{\kappa _m }}{2})m-ig_{ma} a-ig_{mb}(b + b^\dag  )m \\
 &- i{\rm{K}}_m (2m^\dag  m+1)m,
\end{aligned}
\end{eqnarray}
where $\Delta_a =\omega_a -\omega_d$ and $\Delta_m =\omega _m-\omega_d$ are, respectively, the detunings from the microwave pumping field and the cavity photon and magnon modes.
$\Delta _p=\omega _p -\omega _d $ is the beat frequency between the microwave pumping and probe fields.
${\kappa_a}$, ${\kappa_b}$ and ${\kappa_m}$ are the decay rate of the microwave cavity mode, the vibrational mode, and the Kittel modes, respectively.
The operators of the microwave cavity, vibrational, and magnon modes are reduced to their expectation values in the semiclassical approximation, viz. $ o(t) = \left\langle {\hat o(t)} \right\rangle $, with $ o  = a, b$, or $m$.
Furthermore, the mean-field approximation by factorizing averages is also used, and the quantum noise terms are dropped safely \cite{noise}.
Magnomechanical interactions, including the radiation pressure-like magnetostrictive effect and the magnon Kerr nonlinearity, involve a wealth of nonlinear physics \cite{kerr11,kerr2,kerr3,comb1,comb2,comb3,comb33,comb4,kerr}, such as mechanical bistability \cite{kerr2,kerr3} and magnonic frequency combs \cite{comb1,comb2,comb3,comb33,comb4}.
It is well known that a nonlinear system is often accompanied by chaotic phenomenon when the nonlinear strength reaches the chaotic threshold \cite{chaos1,chaos2,chaos3,chaos4,chaos7,chaos8}.
A very natural question is whether the mechanical effects of magnon, similar to the mechanical effects of light \cite{optomechanics}, can trigger chaotic motion.

In order to facilitate the discussion of the chaotic characteristics of the system, we define the mean value of the operator as $o  = o _r  + io _i$, here $o_r$ and $o_i$ are real numbers.
Using Euler's formula, we can obtain the equation of motion in the absence of imaginary number, as follows
\begin{eqnarray}\label{eqn:3}
\dot {a_r} &=& \Delta _a a_i-\frac{{\kappa_a}}{2}a_r+ g_{ma} m_i+\sqrt{{\kappa _{\rm{1}}}} \big[{\varepsilon _d}\sin {\varphi _d}+{\varepsilon _p}\sin(\varpi)\big], \nonumber \\
\dot {a_i} &=& -\Delta _a a_r-\frac{{\kappa_a}}{2}a_i -g_{ma}m_r-\sqrt {{\kappa _{\rm{1}}}}\big[{\varepsilon _d}\cos {\varphi_d}+{\varepsilon _p}\cos(\varpi)\big], \nonumber \\
\dot {b_r} &=& \omega _b b_i - \frac{{\kappa_b }}{2}b_r, \\
\dot {b_i} &=& -\omega _b b_r  - \frac{{\kappa_b }}{2}b_i  - g_{mb} (m_r^2  + m_i^2 ), \nonumber \\
\dot {m_r} &=& g_{ma} a_i- \frac{{\kappa _m }}{2}m_r + {\aleph}m_i,  \nonumber \\
\dot {m_i} &=& - g_{ma} a_r - \frac{{\kappa _m }}{2}m_i - {\aleph}m_r, \nonumber
\end{eqnarray}
here, $\varpi=\Delta_{p}t+\varphi_p$ and $\aleph {\rm{ = }}2g_{mb} b_r  + \Delta _m  + 2{\rm{K}}_m (m_r^2  + m_i^2 ) + {\rm{K}}_m$.
Furthermore, to describe the hypersensitivity of the system to initial conditions (the so-called butterfly effect), a perturbation $\vec \delta  = (\delta a_r ,\delta a_i ,\delta b_r ,\delta b_i ,\delta m_r ,\delta m_i )^\mathrm{T} $ is considered, which characterizes the degree of divergence or convergence of adjacent trajectories in phase space.
The evolution of the perturbation $\vec \delta$, therefore, is derived by linearizing Eqs. (\ref{eqn:3}) as $d\vec \delta /dt = {\rm{M}}\vec \delta $ \cite{chaos1}, with the coefficient matrix
\[
{\rm{M}}=
\begin{pmatrix}
 { - \frac{{\kappa _a }}{2}} & {\Delta _a } & 0 & 0 & 0 & {g_{ma} }  \\
    { - \Delta _a } & { - \frac{{\kappa _a }}{2}} & 0 & 0 & { - g_{ma} } & { 0 }  \\
    0 & 0 & { - \frac{{\kappa _b }}{2}} & {\omega _b } & 0 & 0  \\
    0 & 0 & { - \omega _b } & { - \frac{{\kappa _b }}{2}} & {2g_{mb} m_r } & { - 2g_{mb} m_i }  \\
    0 & {g_{ma} } & {2g_{mb} m_i } & 0 & B_1 & A_1  \\
    { - g_{ma} } & 0 & { - 2g_{mb} m_r } & 0 & A_2 & B_2
\end{pmatrix},
\]
where
\begin{eqnarray}\label{eqn:4}   \nonumber
A_1&=&2g_{mb} b_r + 6{\rm{K}}_m m_i^2  + 2{\rm{K}}_m m_r^2 +\Delta _m +{\rm{K}}_m, \\ \nonumber
A_2&=&-2g_{mb} b_r  - 6{\rm{K}}_m m_r^2  - 2{\rm{K}}_m m_i^2 -\Delta _m -{\rm{K}}_m, \\ \nonumber
B_{1,2}&=&\pm 4{\rm{K}}_m m_r m_i -{\kappa_m}/{2}. \nonumber
\end{eqnarray}\label{eqn:4}
The temporal evolution of adjacent trajectories in phase space $\delta I_m$ ($\delta I_m  = \left| {m + \delta m} \right|^2  - I_m$, here, $I_m = |m|^2$ is the intensity of the magnon mode) can be acquired by numerically solving the Eqs. (\ref{eqn:3}) and the perturbation equation $d\vec \delta /dt = {\rm{M}}\vec \delta $ together.
The general solution can be written as $\delta I_m(t) = \delta I_m(0)e^{\lambda_{LE}t}$, and
the logarithmic slope, i.e.,
\begin{eqnarray}\label{eqn:4}
\begin{aligned}
  &\lambda_{LE}=\lim_{t\rightarrow\infty}\lim_{\delta I_m(0)\rightarrow 0}\frac{1}{t}\ln\bigg|\frac{\delta I_m(t)}{\delta I_m(0)}\bigg|.
\end{aligned}
\end{eqnarray}
defines the Lyapunov exponent, which quantifies the chaotic degree of the system and the sensitivity of the system to the initial conditions \cite{chaos1}.
A positive Lyapunov exponent ($\lambda_{LE}>0$) implies divergence and sensitivity to initial conditions. If, conversely, the Lyapunov exponent is negative ($\lambda_{LE}<0$), then the trajectories of two systems with infinitesimally different initial condition will not diverge.
In particular, a zero Lyapunov exponent ($\lambda_{LE}=0$) indicates that the orbits maintain their relative positions and are on a stable attractor \cite{chaos7}.
In what follows, we will discuss in detail the realization of chaotic motion by introducing phase modulation in the case of weak nonlinear magnomechanical interactions.
First of all, for purpose of discussing the phase-dependent effects more convenient, we consider the transformation $\tilde a= ae^{i\varphi_p}\ (\tilde a^\dag=a^\dag e^{-i\varphi_p})$.
Thus, the Hamiltonian of the two-tone microwave drive field in Eq. (\ref{eqn:1}) should be rewritten as $H_{in}=\hbar\sqrt{\kappa _{\rm{1}}} \big\{\big[{\varepsilon _d}{e^{-i\Phi}} +\varepsilon_p e^{-i{\Delta_{p}t}}\big]a^\dag-\rm{H.c.}\big\}$ (in the frame rotating at $\omega_{d}$).
Here, $\Phi$ is the relative phase of the two-tone microwave input field, i.e., $\Phi=\varphi_d-\varphi_p$.
Thereupon, we only need to discuss the dependence of magnomechanical chaos on the relative phase $\Phi$.

\section{Results and discussion}

\begin{figure}[htbp]
\centering
\includegraphics [width=1\linewidth] {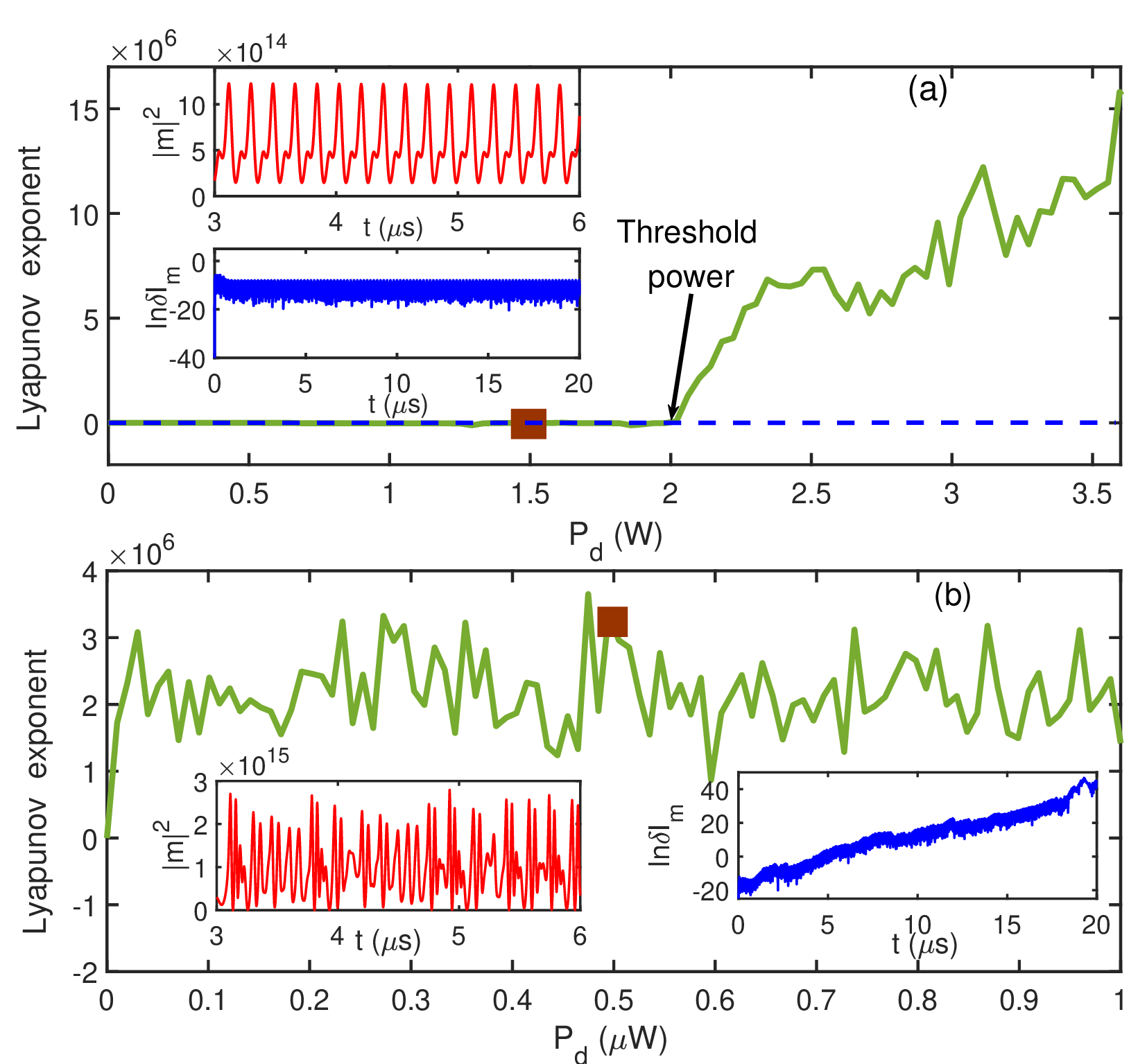}
\caption{(a)-(b) The Lyapunov exponent varies with the microwave driving field power ${\rm{P}}_d$ in the absent and present of the phase modulation.
The inset: the intensity of the magnon mode $|m|^2$ and the perturbation $\ln\delta \rm{I}_m$ vary with time under the microwave driving field power (a) ${\rm{P}}_d = 1.5\ \rm{W}$ and (b) ${\rm{P}}_d = 0.5\ \rm{\mu W}$, respectively.
The microwave drive field power is equal to the probe field power, i.e., ${\rm{P}}_d = {\rm{P}}_p$, and is used throughout the whole work.
The parameters are chosen from the recent experiments \cite{magnomechanics4,kerr3}: $\omega_{d}/{2\pi}=7.645\ \rm{GHz}$, ${\omega_b}/{2\pi}=11.03\ \rm{MHz}$, ${g_{ma}}/{2\pi}=7.37\ \rm{MHz}$, ${g_{mb}}/{2\pi}=10\ \rm{mHz}$, ${\rm{K}}_m/{2\pi}=-6.5\ \rm{nHz}$, ${\kappa_a}/{2\pi}=2.78\ \rm{MHz}$, ${\kappa_b}/{2\pi}=550\ \rm{Hz}$, ${\kappa_1}/{2\pi}=0.22\ \rm{MHz}$, ${\kappa_m}/{2\pi}=2.2\ \rm{MHz}$, and ${\Delta_a}={\Delta_m}=\Delta_{p}={\omega_b}$.
The initial conditions are chosen as $(\vec o)^{\mathrm{T}}$ = $(a_r, a_i, b_r, b_i, m_r, m_i )^{\mathrm{T}}$ = $(0,0,0,0,0,0)^{\mathrm{T}}$, $(\vec \delta)^{\mathrm{T}}$ = $(\delta {a_r }, \delta {a_i }, \delta {b_r }, \delta {b_i }, \delta {m_r }, \delta {m_i })^{\mathrm{T}}$ = $(10^{-10}, 10^{-10}, 10^{-10}, 10^{ - 10}, 10^{ - 10}, 10^{ - 10} )^{\mathrm{T}}$, respectively.}
\label{fig:2}
\end{figure}

Figure (\ref{fig:2}) shows the Lyapunov exponent varies with the microwave driving field power ${\rm{P}}_d$ in the presence and absence of phase mediation.
To be specific, when there is not phase modulated, i. e., the initial phase of the two-tone microwave input field is zero, as shown in Fig. \ref{fig:2}(a).
We can see obviously that the weak nonlinear magnetostrictive interaction of magnons will present a challenge for generating magnomechanical chaos.
For example, when the microwave drive field power is up to ${\rm{P}}_d = 1.5$ W, the Lyapunov exponent is 0 [brown dot in Fig. \ref{fig:2}(a)].
The oscillation of the magnons in the temporal domain is periodic, as the inset shown in Fig. \ref{fig:2}(a), and the flat evolution of $\ln\delta \rm{I}_m$ indicates that the trajectories of nearby points in phase space with infinitesimally disturbance will not diverge.
In this case, we have to continue to increase the microwave drive field power to enhance the nonlinear response of the system.
Understandably, as the microwave drive power increases, the nonlinear response of the system also enhances.
When the nonlinear intensity reaches the chaos threshold, the evolution of the system will change from an ordered state to a chaotic state \cite{chaos2}.
The numerical simulation results show that the threshold of the driving field power required to generate magnomechanical chaos is ${\rm{P}}_d \sim 2.0$ W [shown in Fig. \ref{fig:2}(a)].
The excessive driving power, disadvantageously, will cause significant thermal noise that can't be ignored. Furthermore, when the system temperature is higher than the Curie temperature of YIG sphere, the ferromagnetism and quantum coherence of YIG sphere will disappear \cite{YIG}.
Besides, under high input power, many other higher order terms may become too important to be ignored, such as the Holstein-Primakoff approximation will no longer apply, and these inevitable effects, undoubtedly, will make the system too complicated to research.
Therefore, it is of great significance to reduce the threshold power of magnomechanical chaos.
Advantageously, we find that the chaos threshold can be greatly reduced by introducing phase modulation.
When the relative phase of the two-tone microwave input field $\Phi = 0.4 \pi$, as shown in Fig. \ref{fig:2}(b), a positive Lyapunov exponent can be obtained even if the driving field power is reduced to the magnitude of microwatts (six orders of magnitude less than the case without the phase modulation).
Take one instance, when the power of the microwave driving field ${\rm{P}}_d =0.5 \mu\rm{W}$ [brown dot in Fig. \ref{fig:2}(b)], an aperiodic oscillation of magnons appears, and the calculated exponential divergence of $\delta{I_m}$ indicates the chaotic regime in which initially nearby points in phase space evolve into completely different states separating, as shown in the inset in Fig. \ref{fig:2}(b).
From the above discussion, we can see that in addition to the driving field power, the phase of the microwave driving field plays a crucial role in the chaotic behavior of the cavity magnomechanical system.

\begin{figure}[htb]
\centering
\includegraphics [width=1\linewidth] {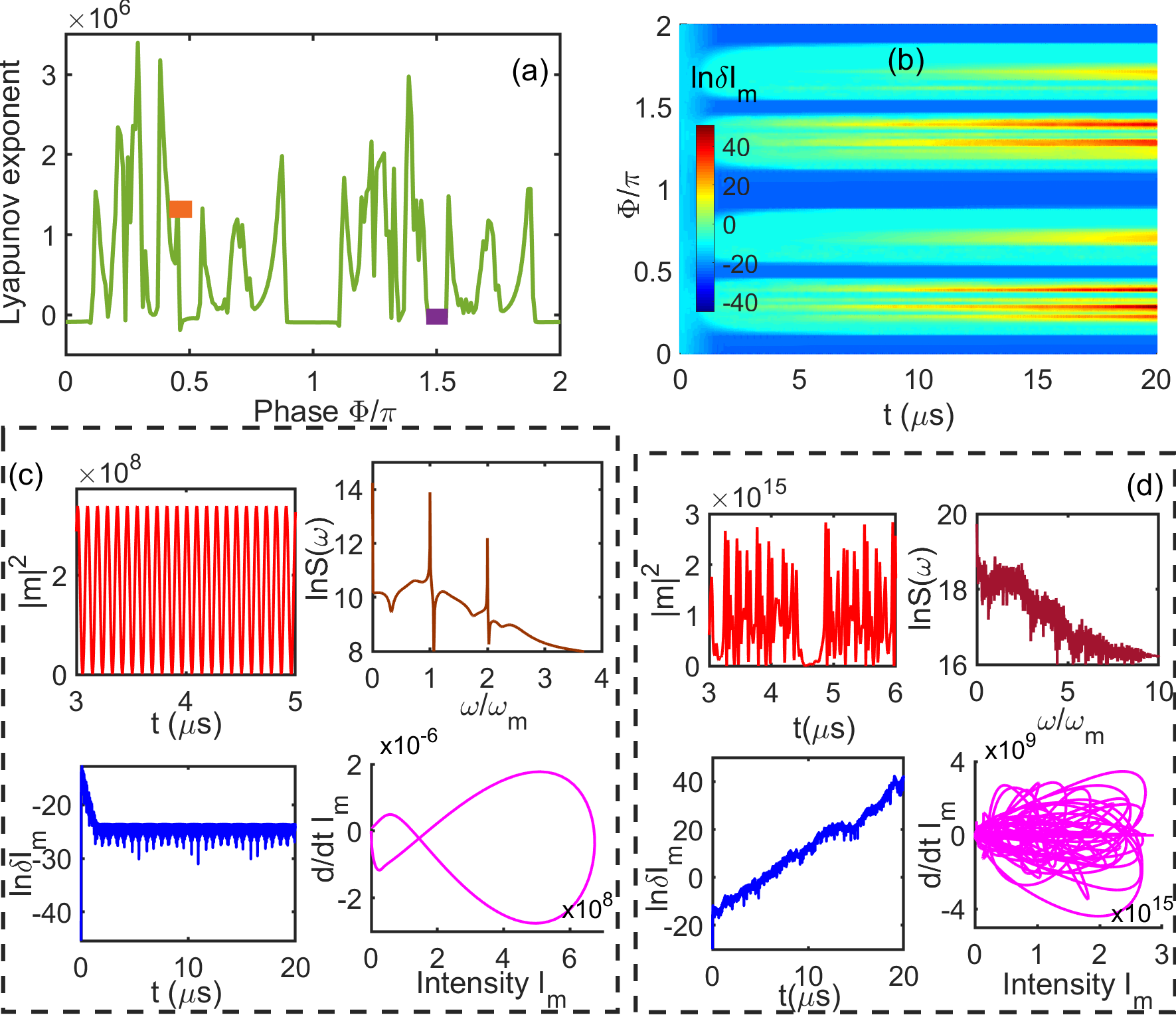}
\caption{(a) The Lyapunov exponent varies with the relative phase of the two-tone microwave input field $\Phi$.
(b) The contour plot of the perturbation $\ln\delta \rm{I}_m$ vary with time t$(\mu s)$ and the relative phase $\Phi$.
The intensity of the magnon mode $|m|^2$ and the perturbation $\ln\delta \rm{I}_m$ vary with time under the different relative phase $\Phi/\pi = 1.5$ in (c) and $\Phi/\pi = 0.4$ in (d), respectively.
Correspondingly, the sideband spectra and the phase-space dynamical trajectories of the magnon are also plotted under the different relative phase $\Phi$. The microwave driving field power is ${\rm{P}}_d = 1 \rm{\mu W}$, and the other parameters are the same as those in Fig. \ref{fig:2}.}
\label{fig:3}
\end{figure}

To further explore the high dependence of the magnomechanical chaotic motion on the phase modulation, the Lyapunov exponent varies with the relative phase of the two-tone microwave input field $\Phi$ has been plotted in Fig. \ref{fig:3}(a).
As the relative phase varies in the range of $0-2\pi$, the Lyapunov exponent alternates between positive and zero, that is, the chaotic oscillation of magnons turns up in some phase regions, and other regions are non-chaotic, including periodic oscillation and period-doubling bifurcation \cite{chaos1,chaos2,chaos3}.
Numerical calculation of the perturbation $\ln\delta \rm{I}_m$ varies the relative phase of the two-tone microwave input field $\Phi$ in the temporal domain [shown in Fig. \ref{fig:3}(b)] confirms these results.
We can clearly see that the perturbation $\ln\delta \rm{I}_m$ changes with the variation of the relative phase, and there are several obvious flat evolution and exponential divergence of $\ln\delta \rm{I}_m$ in Fig. \ref{fig:3}(b), which shows excellent agreement with Fig. \ref{fig:3}(a).
Furthermore, in order to describe the nonlinear dynamic behavior of the system more comprehensively, the intensity of the magnon mode $|m|^2$, the perturbation $\ln\delta \rm{I}_m$, the sideband spectra, as well as the phase-space dynamical trajectories of the magnon have been discussed under the different relative phase of the two-tone microwave input field.
Two kinds of specific situations are analysed in detail.
When the relative phase $\Phi/\pi = 1.5$ [brown dot in Fig. \ref{fig:3}(a)], the Lyapunov exponent is zero, which means that the evolution of the magnons appears period-doubling bifurcation \cite{chaos2}.
In the temporal domain, the non-monochromatic magnonic oscillation $|m|^{2}$ and the flat evolution of the perturbation $\delta \rm{I}_m$ well demonstrate the period-doubling bifurcation process.
In the frequency domain, the spectrum of the magnonic dynamics $S(\omega)$ ($\omega$ is the spectroscopy frequency), obtained by performing the fast Fourier transform of the time series, also conforms to this dynamic behavior.
In addition, as shown in Fig. \ref{fig:3}(c), the dynamical trajectory of magnon evolution in phase space under infinitesimally initial perturbation will finally oscillate in the limited circles.
In another case, when the relative phase of the two-tone microwave input field $\Phi/\pi = 0.4$
[purple dot in Fig. \ref{fig:3}(a)], a positive Lyapunov exponent has been obtained, which means that the system is extremely sensitive to slight changes in the initial conditions.
The aperiodic oscillation of the magnon intensity ${I}_m$ and the continuum sideband spectra well verify this chaotic behaviour \cite{chaos7}.
Moreover, the perturbation $\delta \rm{I}_m$ diverge exponentially, implying that the system is extremely sensitive to the initial condition, which is one of the basic characteristics of chaotic motion \cite{chaos10}.
The evolution of initial nearby trajectory in phase space, as shown in Fig. \ref{fig:3}(d), becomes unpredictable and random.
From the above discussion, we can see that the cavity magnomechanical chaos can be easily realized by phase modulation and the transition from order to chaos can be regulated, which is of great significance to the study of chaotic motion and its regulation in the cavity magnomechanics.

\begin{figure}[htbp]
\centering
\includegraphics [width=1\linewidth] {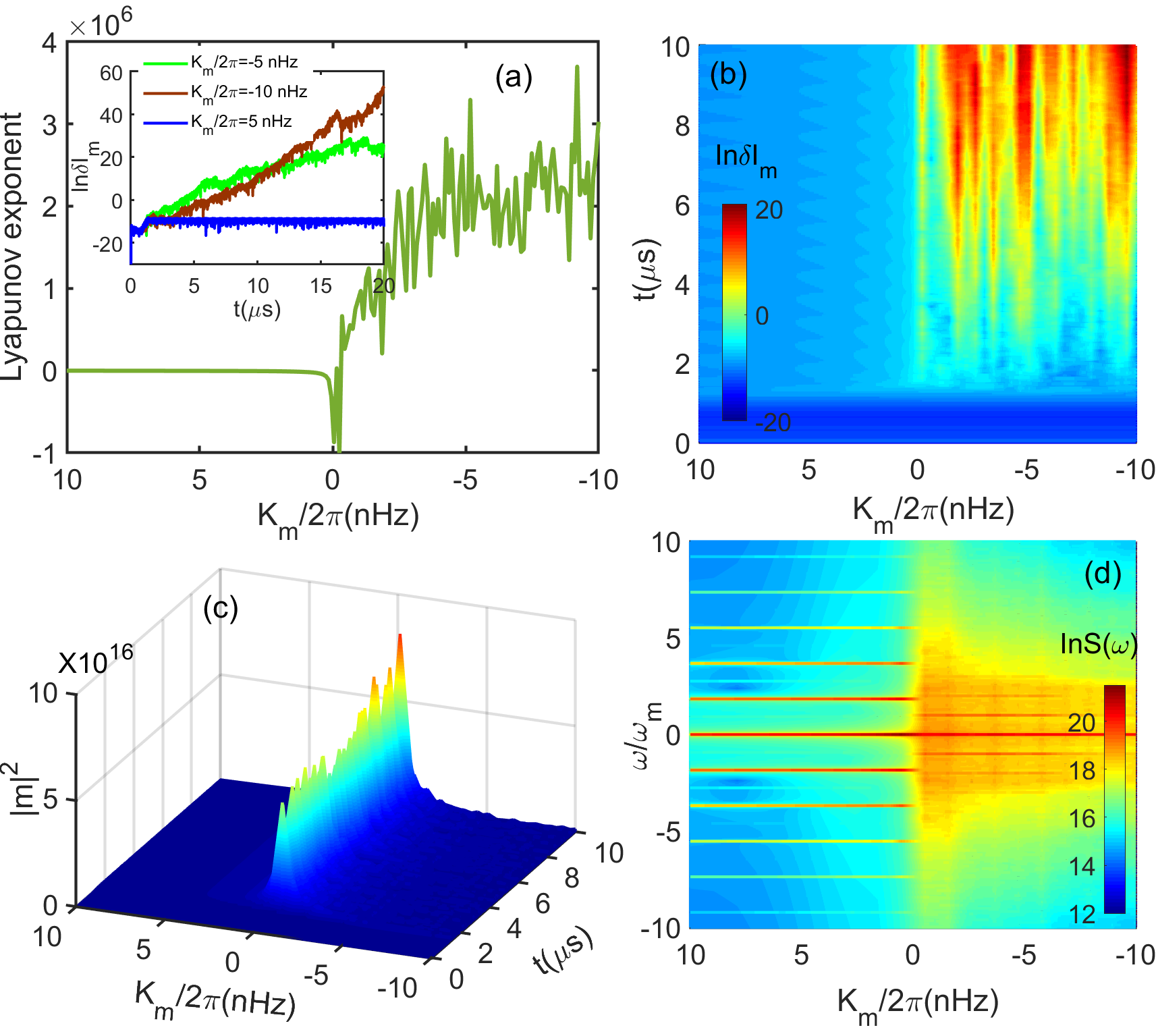}
\caption{(a) The Lyapunov exponent varies with the magnon Kerr coefficient ${\rm{K}}_{\rm{m}}$.
 The inset: the perturbation $\ln\delta \rm{I}_m$ under different magnon Kerr coefficients ${\rm{K}}_{\rm{m}}$.
(b) The contour plot of the perturbation $\ln\delta \rm{I}_m$ vary with time t$(\mu s)$ and the magnon Kerr coefficient ${\rm{K}}_{\rm{m}}$.
(c) The surface plot of the magnonic evolution $|m|^2$ with different magnon Kerr coefficient ${\rm{K}}_{\rm{m}}$.
(d) The contour plot of the magnonic sideband spectra with different magnon Kerr coefficient ${\rm{K}}_{\rm{m}}$.
The microwave driving field power is ${\rm{P}}_d = 1 \rm{\mu W}$ and the relative phase of the two-tone microwave input field $\Phi/\pi = 0.4$. The other parameters are the same as those in Fig. \ref{fig:2}.}
\label{fig:4}
\end{figure}

Up to now, we have shown the generation and manipulation of the cavity magnomechanical chaos induced by phase modulation.
It can be seen from Eq. (\ref{eqn:1}) that the system nonlinearity is derived from two different kinds of nonlinearities, namely, the radiation-pressure-like magnetostrictive interaction and the magnon Kerr nonlinearity \cite{Hybrid,S. Zheng2023}.
Notably, the Kerr coefficient is inversely proportional to the volume $\rm{V_{m}}$ of the YIG sphere, i.e., $\rm{K_{m}} \propto \rm{V_{m}^{-1}}$, and thus, the Kerr effect of magnons can become important for a small YIG sphere.
Furthermore, the Kerr coefficient becomes positive or negative when the crystallographic axis [100] or [110] of the YIG is aligned along the static field $B_{o}$ \cite{kerr}.
Therefore, it is necessary to discuss the influence of the magnon Kerr effect on chaotic dynamics.
To this aim, numerical calculation of the Lyapunov exponent varying with the magnon Kerr coefficient ${\rm{K}}_{\rm{m}}/2\pi$ from 10 nHz to -10 nHz has been shown in Fig. \ref{fig:4}(a).
Intriguingly, when the magnon Kerr coefficient changes from 0 to 10 nHz, i.e., the [110] axis of the YIG sphere is parallel to the static magnetic field, the Lyapunov exponent is always zero.
This implies that the trajectories of two adjacent points with infinitely small initial conditions will not diverge, indicating that the system is in a non-chaotic regime.
However, when the magnetic field direction is changed so that the [100] axis of the YIG sphere is parallel to the static magnetic field, the magnon Kerr coefficient $K_{m}$ is negative.
Under this circumstance, a positive Lyapunov exponent indicates a totally different regime in which initially nearby points in phase space evolve into completely different states.
Moreover, the chaotic degree of the system changes constantly when the magnon Kerr coefficient varies from 0 to -10 nHZ.
More specifically, the temporal evolution of the perturbation $\ln\delta \rm{I}_m$ with different magnon Kerr coefficient ${\rm{K}}_{\rm{m}}/2\pi$ = 5, -5, and -10 nHz are shown by the blue, green, and brown lines in the illustration in Fig. \ref{fig:4}(a), respectively.
It is worth noting that when the Kerr coefficient $\rm{K_{m}}/2\pi=0$, the Lyapunov exponent is negative, indicating that the system is in a periodic state.
This reveals that the appearance of magnomechanical chaos is the result of the combined effect of magnetostrictive interaction and magnon Kerr nonlinearity.
Furthermore, a high dependence of the perturbation evolution on the magnon Kerr coefficient is observed in Fig. \ref{fig:4}(b).
Among them, the flat evolution of $\ln\delta \rm{I}_m$ and the exponential divergence of $\delta \rm{I}_m$ are, respectively, observed in the region of ${\rm{K}}_{\rm{m}}/2\pi \in (10, 0)$ nHz and ${\rm{K}}_{\rm{m}}/2\pi \in (0, -10)$ nHz, which show an excellent agreement with the result in Fig. \ref{fig:4}(a).
Likewise, the magnonic evolution and the magnonic sideband spectrum with different magnon Kerr coefficients are also investigated for the sake of verifying the influence of the magnon Kerr effect on chaotic dynamics.
The the periodic and aperiodic oscillations of the mangons, as well as the separated and continuous sideband spectra, as shown in Figs. \ref{fig:4}(c) and (d), correspond one-to-one with the results in Fig. \ref{fig:4}(a).

Finally, we give some discussion on the feasibility of the experimental realization of the cavity magnomechanical chaos.
First, the present system is simple and has high feasibility in experimental implementation.
The magnetostrictive interaction and the magnon Kerr effect have been experimentally demonstrated, and the simulation parameters used in this work are chosen from the recent experiments \cite{magnomechanics4,kerr3}.
Second, for a YIG sphere with the diameter 0.28-mm, a negative magnon Kerr nonlinear coefficient can be yielded ${\rm{K}}_{\rm{m}}/2\pi \approx$ -6.5 nHz when the [110] axis of the YIG sphere aligned parallel to the static magnetic field \cite{kerr3}, which is well above the threshold for triggering chaotic motion required for our theoretical calculations.
Furthermore, the magnon Kerr coefficient can be further strengthened by reducing the volume $V_{m}$ of the YIG sphere \cite{kerr2}.
On the other hand, for the experimental detection of the magnomechanical chaos, the spectral information of the magnon can be conveniently readout through the microwave photons using a three-dimensional copper cavity, as the experiments \cite{magnomechanics4} have done.
Third, Kerr-modified magnomechanical chaos may also hold for other magnon-coupled systems because magnon possess excellent compatibility with other quasiparticles (for example, photons and qubits).
Finally, with the advancement of nanoprocessing technology, YIG spheres can be easily integrated with on-chip devices, and magnomechanical chaos may find potential applications in secure communication based on magnetic devices.

\section{CONCLUSION}
To conclude, nonlinear chaotic dynamics in the cavity magnomechanical system is discussed in detail.
Using the same parameters as the recent cavity magnomechanical experiments, we identify that the outstanding challenge that weak nonlinear magnomechanical interaction cannot trigger chaotic motion can be effectively solved by introducing phase modulation.
The results indicate that the relative phase of the two-tone input field has a significant affect on the dynamic of the system, thereby inducing the appearance of ultra-low threshold chaotic motion.
Furthermore, the chaotic behavior exhibits a high dependence on the magnon Kerr nonlinearity, which reminds us of the possibility that the "on" and "off" of chaotic motion can be realized by adjusting the direction of the applied magnetic field.
Beyond their fundamental scientific significance, the investigation of magnomechanical chaos will deepen our understanding of nonlinear magnomechanical interaction and can find general relevance to other nonlinear systems based on magnonics.

$\nonumber\\$

\textbf{Author contribution statement}:
Jiao Peng: Carried out the calculations, Wrote the main manuscripttext, Prepared all figures, Reviewed the manuscript, Writing of the manuscript. Zeng-Xing Liu: Participated in the discussions, Reviewed the manuscript, Contributed to the interpretation of the work, Writing of the manuscript. Ya-Fei Yu: Participated in the discussions, Reviewed the manuscript. Hao Xiong: Participated in the discussions, Reviewed the manuscript.

$\nonumber\\$

\textbf{Data Availability Statement}:
Data underlying the results presented in this paper are not publicly available at this time but may be obtained from the corresponding author upon reasonable request.

$\nonumber\\$

\textbf{Conflict of Interest}:
The authors declare no conflicts of interest.

$\nonumber\\$

\textbf{Acknowledgments}:
This work was supported by the National Science Foundation (NSF) of China (Grants No. 12105047), Guangdong Basic and Applied Basic Research Foundation (Grant No. 2022A1515010446), Guangdong Provincial Quantum Science Strategic Initiative) (GDZX2305001), Guangdong Provincial Quantum Science Strategic Initiative) (GDZX2303007).

\end{document}